\documentclass[%
reprint,
amsmath,amssymb,
prb,
]{revtex4-1}
\usepackage{graphicx}% Include figure files
\usepackage{dcolumn}% Align table columns on decimal point
\usepackage{bm}% bold math
\usepackage[bookmarks]{hyperref}% add hypertext capabilities
\usepackage[version=3]{mhchem}
\usepackage[all]{hypcap}
\usepackage{float}

\begin{document}
	
	\preprint{APS/123-QED}
	
	\title{Remarkably enhanced Curie temperature in monolayer  CrI$_3 $ by hydrogen and oxygen adsorption: A first-principles calculations}% Force line breaks with \\

	\author{Maedeh Rassekh$^{1,3}$}\email{rasekh.maede@gmail.com}
	\author{Junjie He$^{2}$}
	\author{Saber Farjami Shayesteh$^{3}$}
	\author{Juan Jose Palacios $^{1,4}$}\email{juanjose.palacios@uam.es}
	
	\affiliation{$^{1}$Departamento de F\' isica de la Materia Condensada, Universidad Aut\' onoma de Madrid, 28049 
		Madrid, Spain.\\
		$^{2}$Bremen Center for Computational Materials Science, University of Bremen, Am Fallturm 1, 28359 Bremen, Germany\\
		$^{3}$Department of Physics, University of Guilan, 41335-1914 Rasht, Iran \\
		$^{4}$ Instituto Nicol\'as Cabrera (INC) and Condensed Matter Physics Center (IFIMAC), 
		Universidad Autónoma de Madrid, 28049 Madrid, Spain.}
	
	\begin{abstract}
		Two-dimensional (2D) materials unique properties and their promising applications in post-silicon microelectronics have attracted significant attention in the past decade. Recently, ferromagnetic order with out-of-plane easy axis in a monolayer of  CrI$_3 $ has been observed and reported, with a Curie temperature of 45 Kelvin. Here we study, using density functional theory (DFT) calculations, how hydrogen and oxygen adsorption affects the structural, electronic, and magnetic properties of a  CrI$_3 $ monolayer. Our results show that while the structure remains almost unchanged by the adsorption of hydrogen, adsorption of oxygen completely distorts it. We have also found that both the adsorption of hydrogen and oxygen atoms significantly influences the electronic and magnetic properties of the monolayer. While hydrogen quenches the magnetic moments of Cr atoms, oxygen introduces an impurity band in the gap. Interestingly, we find a strong enhancement of the Curie temperature by full hydrogenation, while the results are not conclusive for O. This result suggests a simple and effective approach to manipulate the electronic and magnetic properties of 2D magnets  for spintronics applications.
		
	\end{abstract}
	
	
	\maketitle
	
	%\tableofcontents
	
	\section{Introduction}
	
	Since the discovery of graphene in 2004, there is a tremendous interest in experimental and theoretical studies as well as development of two-dimensional (2D) materials \cite{novoselov2004electric}. Various 2D materials such as 
	hexagonal boron nitride \cite{dean2010boron}, silicene  \cite{cahangirov2009two}, and transition metal dichalcogenides
	\cite{mak2010atomically} have been identified as important candidates for next generation electronic and optoelectronic devices due to their unique properties. Although almost all types of electronic behavior such as metallic,
	semiconducting, insulating (topological and trivial), and superconducting have been observed and studied, magnetism has remained elusive until very recently. The discovery of the first truly 2D magnet,  CrI$_3 $, has completed the family of 2D materials. A single atomic layer of   CrI$_3 $ is magnetic and its Curie temperature is about 45$^{\circ}$K, what can be considered a relatively high temperature \cite{huang2017layer}. Still far from room temperature, we can hope that reaching this goal will not be impossible, which would be of unlimited potential in nanoscale magnetic memory devices and spintronics where spin-based logic operations instead of charge-based ones could guarantee low power consumption and extended device applications \cite{wang2019theoretical}. Technically, one typically needs two magnetic metals or half-metal electrodes and a non-magnetic semiconductor in between or, alternatively, a magnetic semiconductor between non-magnetic electrodes to have a spintronic device.  The recent reports of magnetic order in different 2D crystals such 
	as $ Fe_3GeTe_3 $ \cite{fei2018two}, $ VSe_2 $ \cite{bonilla2018strong}, $ Cr_2Ge_2Te_6 $ \cite{gong2017discovery}, $ FePS_3 $ 
	\cite{lee2016ising,wang2016raman}, and MXenes \cite{he2019cr} provide us with an exciting new platform in research of 2D materials and exploration of fundamental theory of magnetism.
	
	Chromium trihalides,  CrX$_3 $ (X = Cl, Br, I), are a class of van der Waals bonded, layered semiconductors. 
	These compounds are magnetic and have been known for many decades \cite{mcguire2015coupling}. Weak van der Waals (vdW) 
	interaction between stacking layers in this kind of materials offer the possibility for exfoliation into 
	single-layer nanosheets \cite{wang2016doping} and, remarkably, their intrinsic ferromagnetism can persist down to few 
	layers and even single layer. Among this family, the exceptional ferromagnetic behavior and the highest magnetic 
	ordering temperature of  CrI$_3 $ can be ascribed to strong covalent interaction between the Cr atom and 
	its nearest-neighbor I atoms. Owing to its electronic band gap, CrI$_3 $ monolayer is a promising material 
	for spintronics applications. This is remarkable because, typically, ferromagnets tend to be quite conductive, 
	while antiferromagnets/ferrimagnets are mostly insulators or semiconductors \cite{gao2019ferromagnetism}. 
	
	All this has prompted a number of studies  \cite{frisk2018magnetic,klein2018probing,jiang2018spin}. 
	For example, based on density functional theory (DFT) calculations, Jiang et al. \cite{jiang2018spin} found that a 
	direct-to-indirect band gap transition can occur by manipulating spin direction of  CrI$_3 $ from out-of-plane to 
	in-plane, which can induce a magnetic field controlled photoluminescence. Jiang et al. \cite{jiang2019first} 
	have studied  CrI$_3 $ zigzag nanoribbons, which are strips of  CrI$_3 $ monolayer. They found that these are also
	ferromagnetic, that the edge states dominate the band structure around the Fermi level, and that this can be 
	tuned by the different edge atomic structures. Moreover, because of the weak van der Waals (vdW) interlayer coupling, 
	designing heterostructures with interesting properties using  CrI$_3 $ and other vdW materials is possible without 
	considering lattice mismatch. As found by Zhang et al. \cite{zhang2018strong}, reducing the distance between 
	graphene and  CrI$_3 $ in a heterostructure to about 3.3 and 2.4 \AA, a Chern insulating state can be acquired. 
	A very strong gap opening of about 150 meV is also found in graphene which is induced by the magnetic proximity 
	of  CrI$_3 $. Song et al. \cite{song2018giant} have reported that atomically thin chromium triiodide (CrI$_3 $) can act as a spin-filter tunnel barrier sandwiched between graphene contacts. Based on their result, increasing the number of CrI$_3 $ layers can enhance tunneling magnetoresistance. The tunneling current ($ I_t $) also has a strong magnetic field dependence which implies a spin-dependent tunneling probability related to the field-dependent magnetic structure of CrI$_3 $. 
	
	Strain engineering is another effective strategy for tuning and controlling 2D materials as well as CrI$ _3 $. The electronic band gap of CrI$ _3 $ along with characteristic ferromagnetism can be controlled by uniaxial and biaxial compressive as well as tensile strains \cite{mukherjee2019strain}.  Wu et al. \cite{wu2019strain} have investigated magnetic and electronic properties of monolayer  CrI$_3 $ under strain. They have found that biaxial strain can induce a ferromagnetic-antiferromagnetic transition. By varying the amount of strain, a transition in its electronic state occurs from magnetic-metal to half-metal and to half-semiconductor. Zheng et al. \cite{zheng2018tunable} results suggest that the out of plane magnetic orientation can flip to in plane by a tensile strain. Webster et al. \cite{webster2018strain} have shown that the Curie temperature of CrI$ _3 $ monolayer is sensitive to the strain, and its highest amount is close to the equilibrium point. As the compressive strain increases the Curie temperature decreases.   
	
	\begin{figure}[t]
		\centering
		\includegraphics[width=8.3cm]{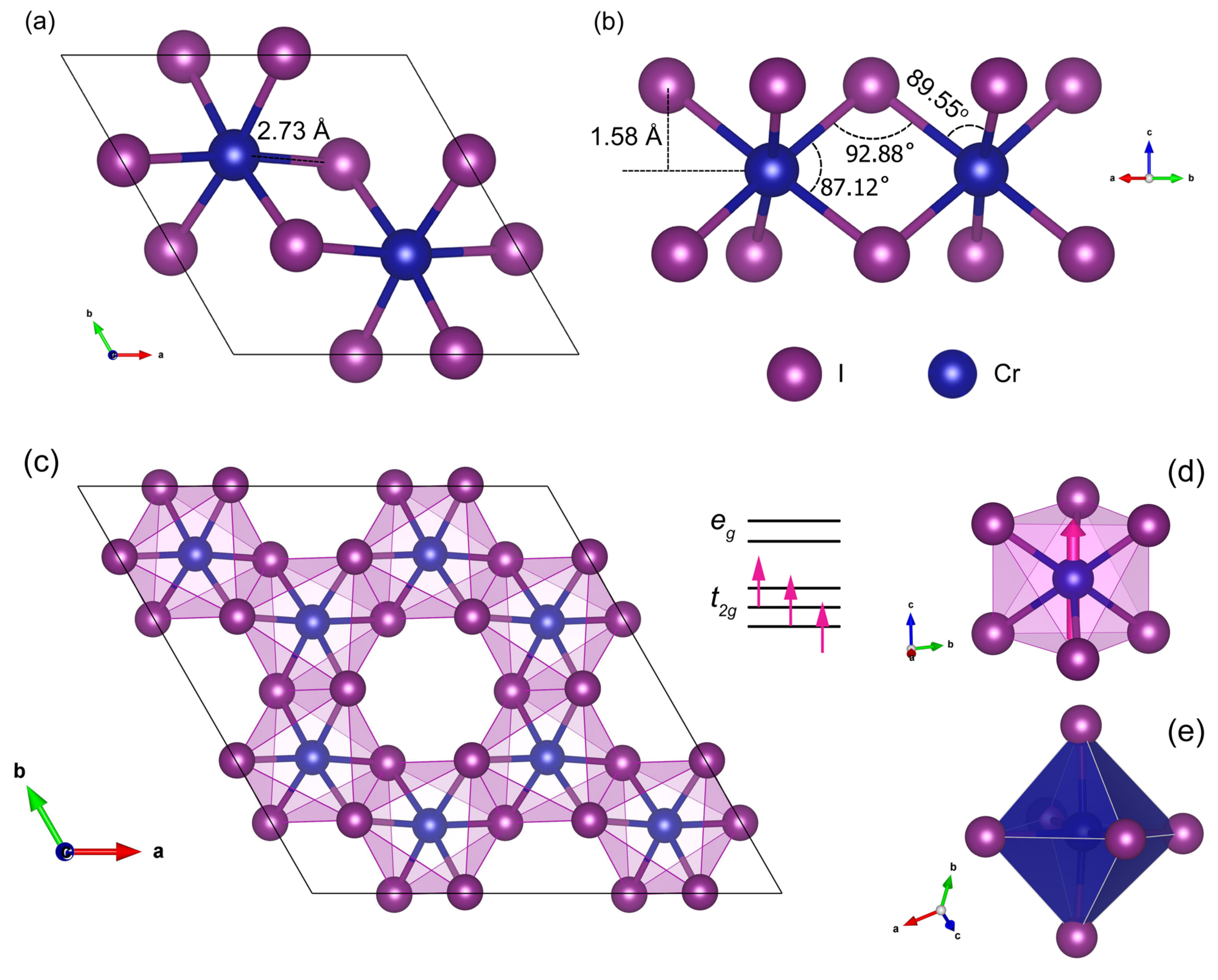}
		\caption{Schematic illustration of the single-layer  CrI$_3 $ unit cell: bond length of Cr-I (a), bond angles and 
			vertical distance between the plane containing I atoms and the plane containing Cr atoms (b), 
			top view of  CrI$_3 $ crystalline structure, showing the honeycomb arrangement of the Cr atoms (c), 
			side view of a single Cr site with an arrow representing its out-of-plane magnetic moment and the 
			splitting of d-levels into a higher energy $e_g$ doublet and a lower energy $ t_{2g} $ triplet in an 
			octagonal environment (d), and the edge-sharing octahedral  Cr$^{3+} $ ion with six  I$^- $ ions (e).}
		\label{fgr:1}
	\end{figure}
	
	Tuning the electronic and magnetic properties of materials has always been a central topic in condensed 
	matter physics and material science. For instance, in this regard, interactions between adatoms and  CrI$_3 $ monolayer are 
	important because the electronic structure of the system can be altered, which could lead to appealing new properties.
	However, the effect of adsorption of atoms on the magnetism of intrinsic 2D magnets is not yet sufficiently understood. In 
	this work, motivated by previous investigations on other 2D materials, we intend to reveal the effect of 
	H and O adsorption on the structural, electronic, and magnetic properties of  CrI$_3 $ monolayer. Our 
	paper is structured in the following manner. Details of computational methodology are described in Sec. 2. 
	We discuss our results in Sec. 3. Finally, conclusions and a summary of our results are given in Sec. 4.
	
	\section{computational Methods and preliminary tests}
	
	We have performed DFT \cite{hohenberg1964inhomogeneous,kohn1965self} calculations using the OpenMX code \cite{OpenMX}, which is based on norm-conserving pseudo-potential method \cite{bachelet1982dr,troullier1991efficient,vanderbilt1990soft,blochl1990generalized,morrison1993nonlocal} with a partial core correction and a linear combinations of pseudo atomic orbitals (LCPAO) as a basis functions \cite{ozaki2003variationally,ozaki2004numerical} here specified by 
	Cr$6.0-s^2p^2d^1 $,  I$7.0-s^2p^2d^1 $,  H$6.0-s^2p^1 $, and  O$6.0-s^2p^2d^1 $. For example, in the case of the Cr 
	atom,  Cr$6.0-s^2p^2d^1 $ means that the cutoff radius is 6.0 Bohr \cite{ozaki2003variationally,ozaki2004numerical}, 
	and that two primitive orbitals for each of s and p components and 
	one primitive orbital for d components are used. The exchange-correlation functional was the spin-polarized 
	GGA-PBE \cite{perdew1996generalized}. All calculations were performed until the change in total energy between two 
	successive iteration steps, converged to less than $ 10^{-6} $ Hartree. For the purpose of the calculation of the 
	density of states (DOS), a Gaussian smearing of the energy levels was applied with standard deviation set to 0.15 eV. 
	A  CrI$_3 $ unit cell with two Cr and six I atoms was initially constructed using experimental values 
	\cite{mcguire2015coupling}. A vacuum spacing of 20 \AA\ in the z direction is used to prevent the interaction of 
	the single-layer  CrI$_3 $ with its periodic images. We converged the total energy with respect to the cutoff energy 
	and kgrid for the unit cell. Based on these convergence tests, a cutoff energy of 220 Ry and $ 11\times11 $ kgrid 
	have been used in all presented results. The unit cell structure was also geometrically optimized. The quasi-Newton 
	eigenvector following method was executed for structural relaxation of all structures until the change in 
	forces between two successive iteration steps were less than $ 1\times10^{-3} $ Hartree/Bohr.
	
	After optimization we get a Cr-I covalent bond length of  2.73 \AA\ with bond angles of 92.88$^{\circ}$, 89.55$^{\circ}$, 
	and 87.12$^{\circ}$. These values are consistent with previous DFT results 
	\cite{wang2016doping,mukherjee2019strain,ghosh2019structural}. As shown in (Fig. \ref{fgr:1}), the Cr layer is 
	sandwiched in between two I layers in the two-dimensional  CrI$_3 $ sheet. The plane containing I atoms is at a 
	vertical distance of 1.58 \AA\ from the plane containing Cr atoms (Fig. \ref{fgr:1}) which form a 
	honeycomb network. The  Cr$^{3+} $ ions are surrounded by 6 nearest-neighbor  I$^- $ ions each of them bonded to 
	two Cr atoms and arranged in an edge sharing octahedral coordination (Fig. \ref{fgr:1}). In an ionic picture, 
	Cr$^{3+} $ has three electrons, with an electronic configuration of $ 3d^34s^0 $. Owing to the octahedral symmetry, 
	the d-orbitals split into three low-lying $ t_{2g} $ and two high-lying $ e_g $ orbitals \cite{skomski2008simple}. 
	Therefore  the three electrons of Cr$^{3+} $ ions in this environment occupy the $ t_{2g} $ with $S = 3/2$. Our 
	spin-polarized DFT calculations reveal an energy difference $ (E_{\rm AFM}-E_{\rm FM}) $ between ferromagnetic (FM) and 
	antiferromagnetic (AFM) order of 15.81 meV/Cr atom, meaning that the FM state of  CrI$_3 $ monolayer is 
	energetically more favorable than the AFM state and in good agreement with previous works \cite{wang2016doping,ghosh2019structural,guo2018half}.
	
\begin{figure}[t]
		\centering
		\includegraphics[width=8.3cm]{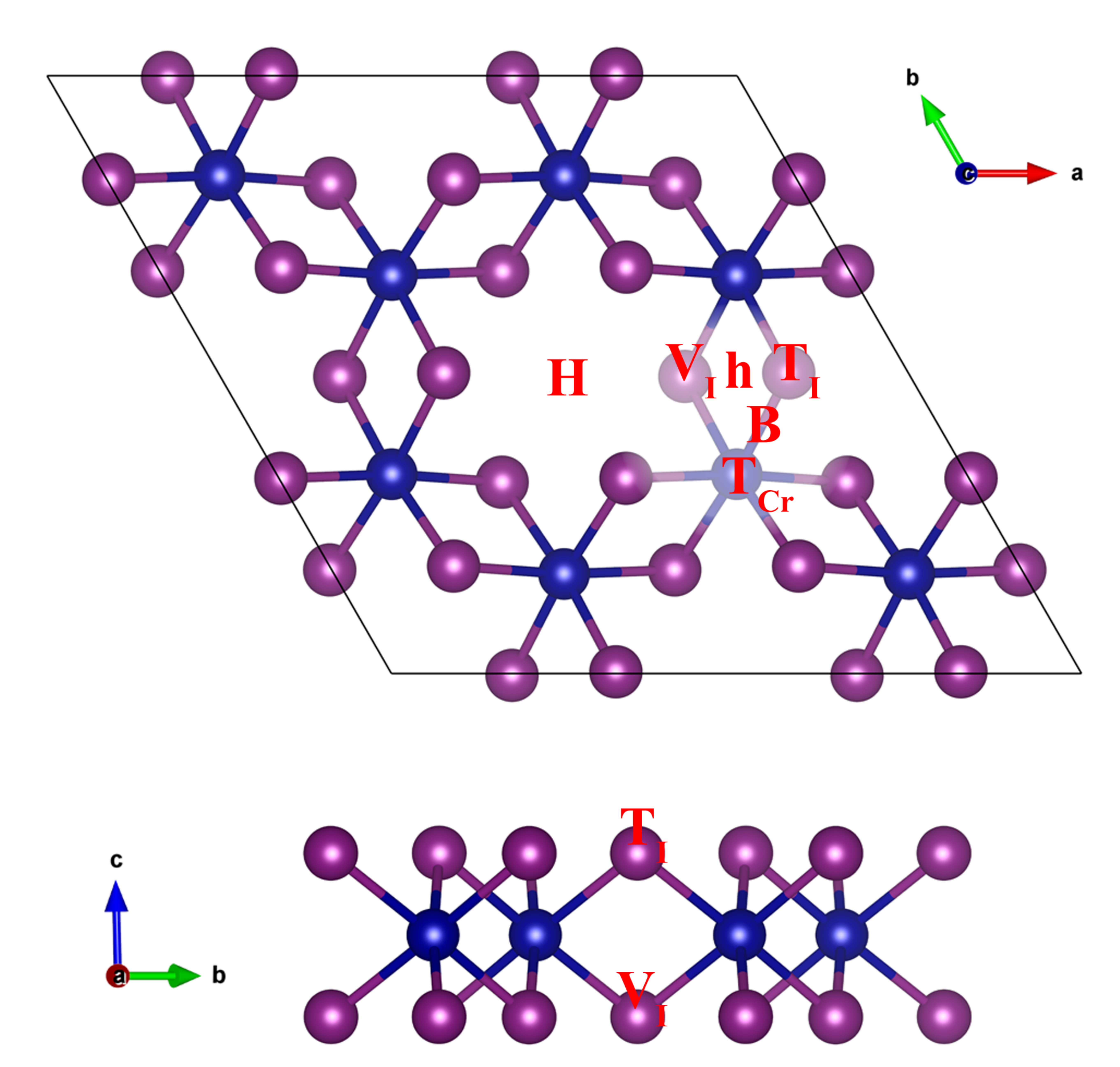}
		\caption{Symbolic presentation of six adsorption sites on CrI$_3 $ lattice.}
		\label{fgr:2}
	\end{figure}

	Hydrogen and O adsorption on CrI$_3 $ monolayer is modeled using single adatom in a $ 2\times2 $ supercell 
	containing 8 Cr and 24 I atoms. For this supercell we re-converged in kgrid and, based on these convergence tests, 
	a $ 7\times7 $ kgrid has been finally used throughout. We have considered adsorption of these atoms at 6 different 
	sites: hollow (H) site at the center of star, hollow (h) site at the center of diamond,  
	bridge (B) site at the midpoint of the Cr-I bond, top ( T$_{\rm I} $) site directly above the upper I atom, 
	valley ( V$_{\rm I} $) site directly above the lower I atom, and top ( T$_{\rm Cr} $) site directly 
	above the Cr atom (Fig. \ref{fgr:2}).
	Unless otherwise mentioned, for each adsorption site, the adatom and  CrI$_3 $ atoms are relaxed in 
	all x, y and z directions. To estimate the adsorption energy of adatoms, the calculations for the isolated adatom, 
	isolated  CrI$_3 $ single layer and  CrI$_3 $ with adatom system are performed in same-sized  CrI$_3 $ supercell.
	
	\section{Results and discussion}
	
\begin{figure}[b!]
		\centering
		\includegraphics[width=8.3cm]{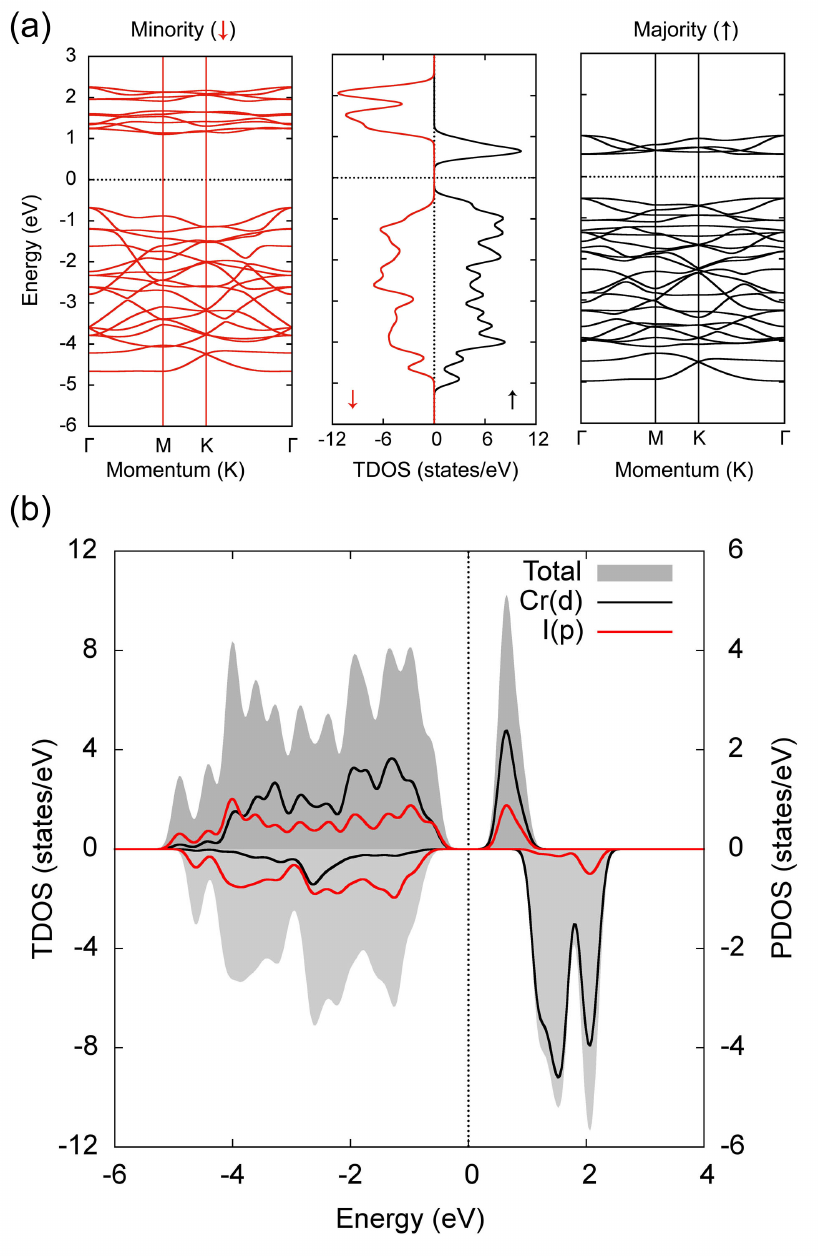}
		\caption{TDOS and band structure of pristine  CrI$_3 $ monolayer (a), and PDOS (b). The positive (negative) 
			values correspond to spin-up (spin-down) states (b).}
		\label{fgr:3}
	\end{figure}

	First we have confirmed that the electronic structure of  CrI$_3 $ monolayer is in good agreement with prior 
	studies  \cite{wang2016doping,lado2017origin,zhang2015robust}. Our calculations show that the single-layer of 
	CrI$_3$ is a ferromagnetic indirect semiconductor with a band gap of about 1.1 eV.  
	The magnetic moment is concentrated mostly at the Cr atoms. 
	The total magnetic moments are 6$ \mu_B $ and 24$ \mu_B $ for the  CrI$_3 $ unit cell and the $ 2\times2 $ supercell, 
	respectively, which is about $ 3\mu_B $ per Cr atom. The total density of states (TDOS) and the band structure 
	for pure  CrI$_3 $ monolayer are plotted in Fig. \ref{fgr:3}.
	\begin{figure*}[ht!]
		\centering
		\includegraphics[width=17.1cm]{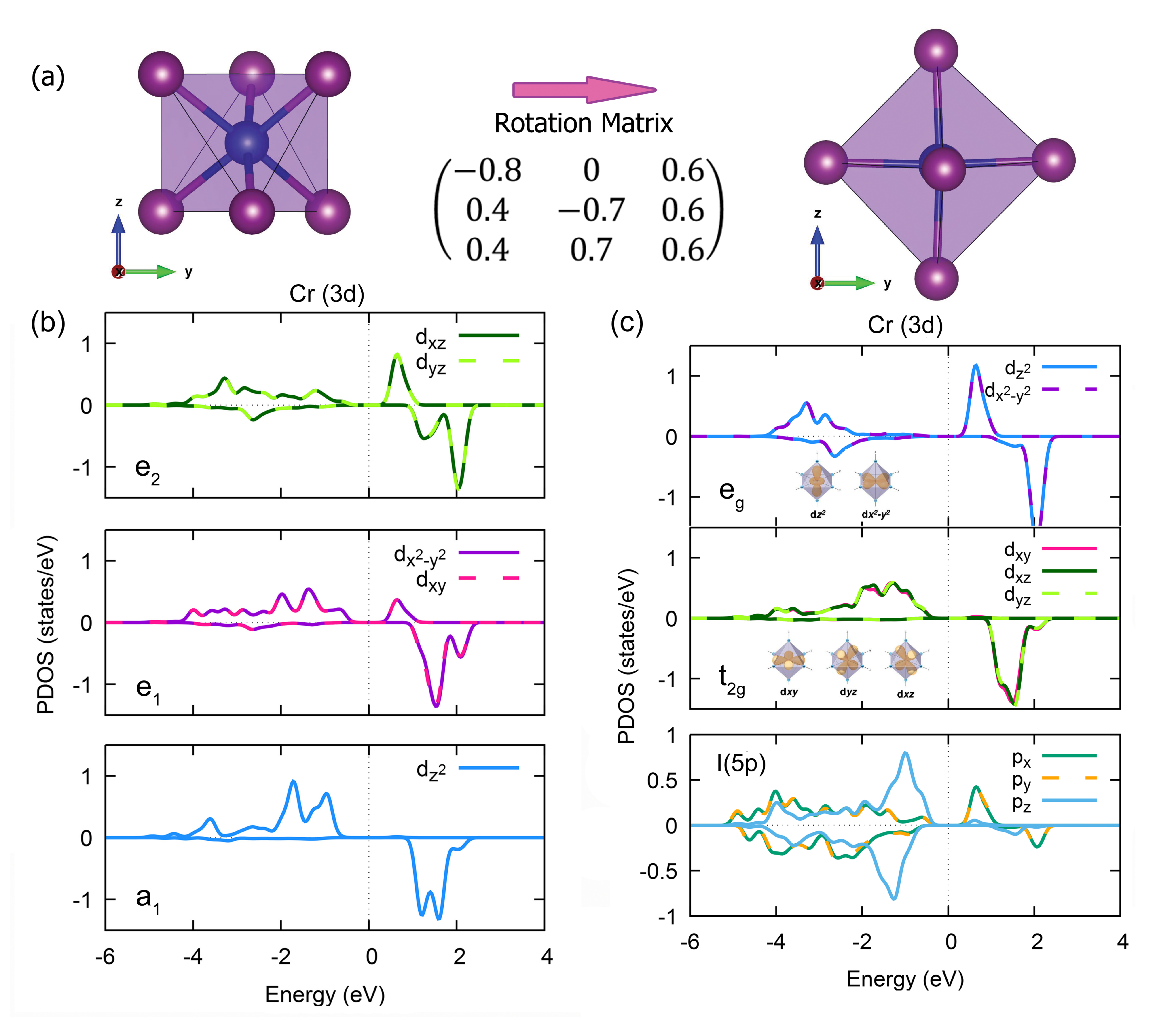}
		\caption{Effect of coordination matching on orbital projections of the  CrI$_3 $ octahedron. 
			Schematic of rotation of  CrI$_3 $ octahedron (a), and corresponding projected DOS in  CrI$_3 $ octahedron
			before (b) and after (c) coordination matching. Before rotation, crystal structure aligned by lattice symmetry: 
			b along y axis, c along z axis, results in unclear projections. After rotation, the Cr-I bonds are exactly 
			aligned parallel to Cartesian axis, allowing DOS orbitally resolved. Blue and purple balls represent 
			Cr and I atoms, respectively.}
		\label{fgr:4}
	\end{figure*}
	The imbalance between the 
	spin-up and spin-down components is evidence of the magnetic nature of the system. The valence band and 
	conduction band edges are formed by spin-up bands, showing a spin-polarized half-semiconductor character. 
	(The band gap for the spin-down channel is about 1.8 eV.)
	This is also reflected in the projected density of states (PDOS), as shown in Fig. \ref{fgr:3}-b, where
	the top of the valence band and bottom of the conduction band are formed 
	mostly by spin-up p orbitals of I atoms and spin-up d orbitals of Cr atoms.
	We can also see the hybridization between Cr-3d and I-5p states. From the PDOS related to the d orbitals 
	of the Cr atom, it is also clear that occupied Cr-3d orbitals are mainly found in the spin-up direction, 
	as we described above using crystal field theory. For more details, we also calculated the 
	PDOS for individual Cr 3d orbitals (see Fig. \ref{fgr:4}-b).
	\begin{table}[b!]
		\small
		\caption{\ Calculated values of various quantities 
			for H and O adsorbed  CrI$_3 $: The most stable site, adsorption energy $ E_a $, the magnetic moment $ \mu $ of the 
			adatom/CrI$_3 $ system, the magnetic moment of an isolated adatom $ \mu_A $, and the charge transfer 
			$ \Delta Q $ between adatom and the CrI$_3 $i monolayer, respectively.}
		\label{tbl:1}
		\begin{tabular*}{0.48\textwidth}{@{\extracolsep{\fill}}cccccc}
			\hline
			Coverage & Site & $ E_a $(eV) & $ \mu (\mu_B) $ & $ \mu_A (\mu_B) $& $ \Delta Q $(e) \\
			\hline
			HCrI$_3 $ ($ 2\times2 $) & $ T_{Cr} $ & 1.02 & 23 & 1  & +0.01\\
			OCrI$_3 $ ($ 2\times2 $) & $ V_I $ & 3.35 & 24 & 2 &  +0.37\\
			\hline
		\end{tabular*}
	\end{table}
	\begin{figure*}[th!]
		\centering
		\includegraphics[width=17.1cm]{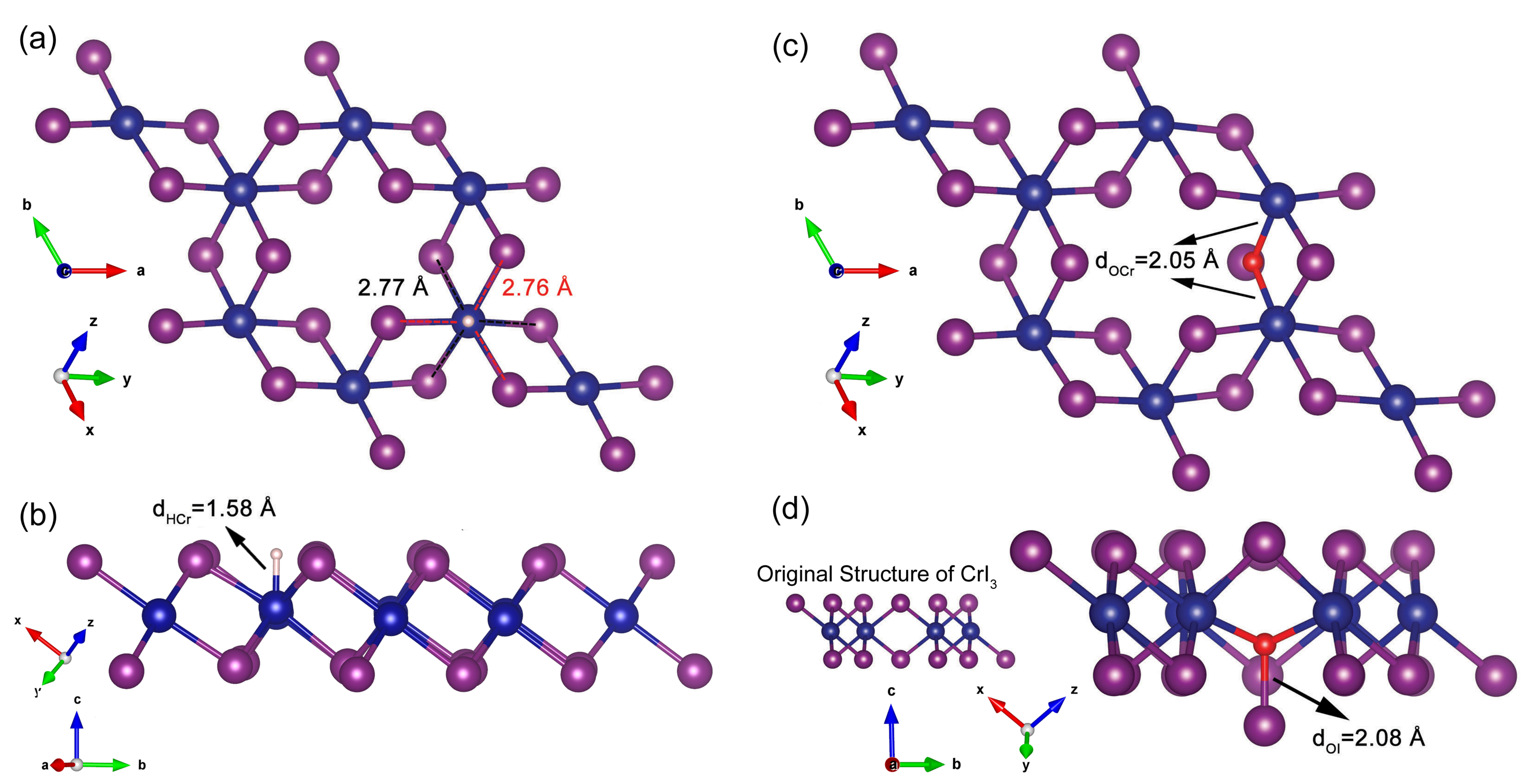}
		\caption{Optimized $ 2\times2 $  CrI$_3 $ supercell with a H atom adsorbed: top view and nearest Cr-I bonds 
			length (a) and H-Cr bond length (b). Optimized $ 2\times2 $  CrI$_3 $ supercell with an O atom attached: 
			top view and nearest O-Cr bonds length (c) and side view and O-I bond length (d). Blue, purple, pink, 
			and red balls represent Cr, I, H and O atoms, respectively.}
		\label{fgr:5}
	\end{figure*}
	\begin{figure}[hb!]
		\centering
		\includegraphics[width=8.3cm]{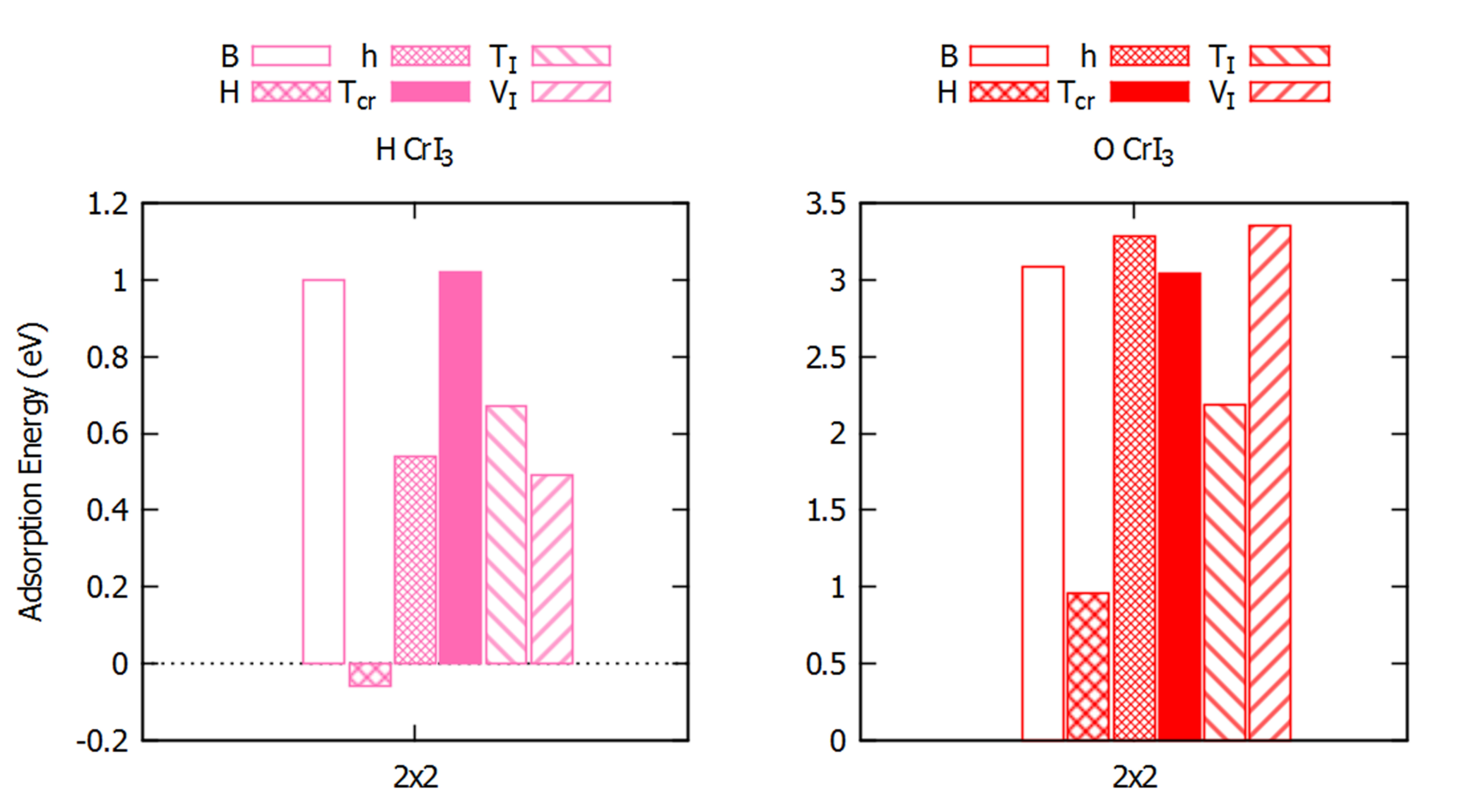}
		\caption{Schematic view of adsorption energy in six adsorption sites:  HCrI$_3 $ (left), and  OCrI$_3 $ 
			(right) systems.}
		\label{fgr:6}
	\end{figure}
	It can be seen that the five Cr 3d states split into three groups ($ a_1=d_{z^2} $, $ e_1=d_{xy} $ and 
	$ d_{x^2-y^2} $, $ e_2=d_{xz} $ and $ d_{yz} $), which is trigonal prismatic crystal field splitting instead of 
	the expected octahedral one. Following Hu et al. \cite{hu2017chemical}, this can be explained by the mismatch
	between the Cartesian coordination with the internal coordination of the octahedral crystal field,
	making the PDOS orbitally unresolved \cite{gao2019ferromagnetism}. To enable an orbitally resolved PDOS that 
	distiguinshes $ t_{2g} $ ($ d_{xy} $, $ d_{xz} $, and $ d_{yz} $) from $ e_g $ ($ d_{x^2-y^2} $, and $ d_{z^2} $) 
	orbitals, a manipulation is needed to align the Cr-I bonds in the octahedrons along x, y, and z axis. 
	For this we first find 
	the proper rotation matrix using the VESTA program \cite{VESTA}. Then we calculated the corresponding rotation angles and
	rotated the  CrI$_3 $ unit cell using Atomsk \cite{Atomsk}. After this change, as we can see in Fig. \ref{fgr:4}-c,
	the PDOS is orbitally resolved and the Cr 3d orbital splits into two groups, $ e_g $ and $ t_{2g} $, which is the expected 
	octahedral splitting. We can also refer this situation to a distorted octahedral crystal field \cite{he2017near}. For the edge-sharing I atoms, 
	their p orbitals split into two $ p_x/p_y $ and $ p_z $ groups. As we can see in Fig. \ref{fgr:4}-c, the stronger 
	$ pd\sigma $ hybridization between in plane $ p_x/p_y $ orbitals and the Cr $ d_{x^2-y^2} $ orbital results
	in lower energies compared to the weaker $ pd\pi $ hybridization between out of plane $ p_z $ and Cr $ d_{xz}/d_{yz} $ 
	orbitals \cite{wang2016doping}. In an ionic picture, we can also see that the I-3p orbitals now are 
	almost completely filled.
	
	Now we turn our attention to the modifications induced by H and O adatoms. 
	The results for the most stable adsorption site for H and O are given in (Table \ref{tbl:1}), in which HCrI$_3 $ 
	and OCrI$_3 $ stand for H- and O-modified systems, respectively. For each system, six adsorption sites (high
	symmetry sites)
	were considered. The relative stabilities of these sites are quantified by their adsorption energy $ E_a $, defined as the
	the energy we need to remove the adatom from the  CrI$_3 $ monolayer: $ E_a=(E_{\rm ad}+E_{\rm CrI_3})-E_{\rm tot} $, 
	where $ E_{\rm ad} $, $ E_{\rm CrI_3} $, and $ E_{\rm tot} $ represent the total energies of a single free adatom, 
	the clean  CrI$_3 $, and the  XCrI$_3 $ system (X = H,O), respectively. 
	For the case with H, the $ T_{Cr} $ site is found to be the most stable structure. As can be seen in Fig. \ref{fgr:6}, 
	all adsorption sites are possible due to their positive absorption energies, except for hollow site.
	For  B ,  T$_{\rm I} $, and  V$_{\rm I} $  sites, we fixed x and y directions of H adatom to prevent it from 
	moving off these sites, and the remaining dimensions are relaxed. For the  OCrI$_3 $ system, the most stable 
	configuration is the  V$_{\rm I} $ site. Optimized $ 2\times2 $  CrI$_3 $ supercell with a H atom adsorbed is shown in 
	Figs. \ref{fgr:5}-a and b. The $ d_{\rm HCr} $ stands for the distances between the H atom and its nearest Cr atom. 
	In this system, the nearest Cr atom is just below the adatom. By the adsorption of the H atom the nearest Cr-I bond 
	length stretches from $ \sim2.73 $ \AA\ to $ \sim2.76 $ \AA\ and 2.77 \AA. While the H atom hardly influences 
	the original  CrI$_3 $ structure, this is greatly distorted by the adsorbed O atom (Fig. \ref{fgr:5}-c and d). In 
	our definition, adsorption energies for all  OCrI$_3 $ structures are positive (see Fig. \ref{fgr:6}), 
	which indicates that attaching oxygen 
	is always energetically more favorable. From Table \ref{tbl:1} it can be seen that the adsorption 
	sites are strongly dependent 
	on the species of adatoms with an adsorption energy for the  OCrI$_3 $ system visibly higher than for  HCrI$_3 $ system.
	
\begin{figure}[t!]
		\centering
		\includegraphics[width=8.3cm]{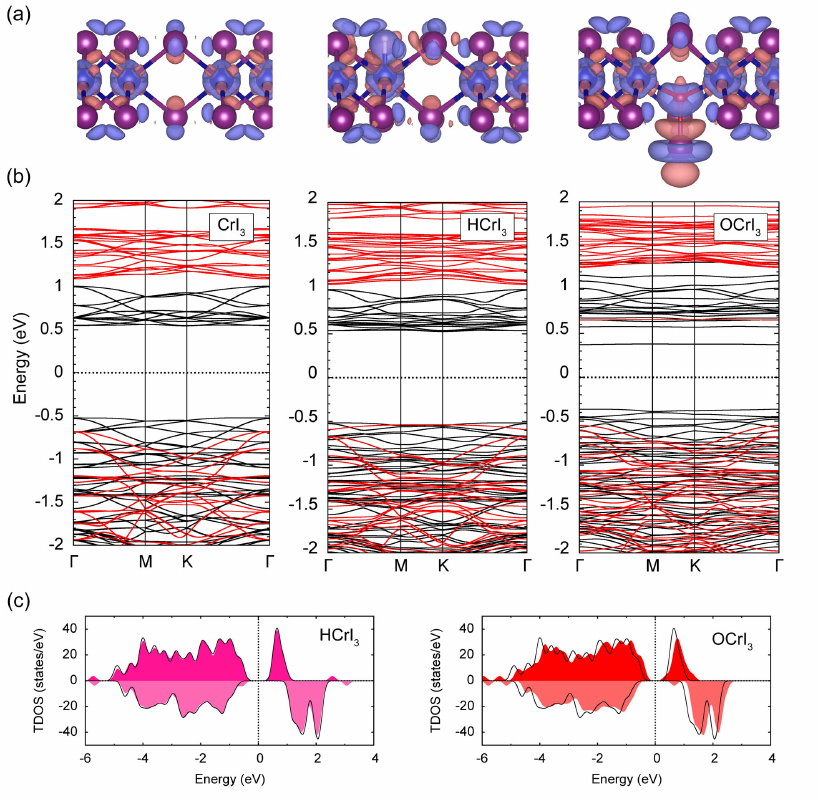}
		\caption{Charge density difference with the isosurface value of 0.005 $ e/Bohr^3 $. The purple and pink represent the charge accumulation and depletion (a), from left to right, the band structures for pure  CrI$_3 $,  HCrI$_3 $, and  OCrI$_3 $ in $ 2\times2 $ systems, respectively. The spin-up and spin-down bands are represented by black and red lines, respectively (b), and TDOS of HCrI$_3 $ and  OCrI$_3 $ systems. The positive (negative) values correspond to spin-up (spin-down) states. Black lines show The TDOS of the pure  CrI$_3 $ (c).}
		\label{fgr:7}
	\end{figure}

	For all calculated TDOS and PDOS we have rotated the structures as we did for the pure  CrI$_3 $ monolayer. 
	We also checked that the results (adsorption energy, magnetic moment, bond length, ...) of the optimized 
	rotated and not-rotated structure are all the same. Again, the only reason for this rotation is to get the 
	appropriate splitting. The TDOS of  HCrI$_3 $ and  OCrI$_3 $ systems are plotted in Fig. \ref{fgr:7}-c
	along with that of the clean  CrI$_3 $ for comparative purposes. As we can see in this figure, H leaves the TDOS 
	of  CrI$_3 $ almost without change, while O introduces visible changes. Most importantly, the latter introduces
	an impurity state spin-up band that appears in the gap (see the band structure in Fig. \ref{fgr:7}-b). 
	
	\begin{figure}[b!]
		\centering
		\includegraphics[width=8.3cm]{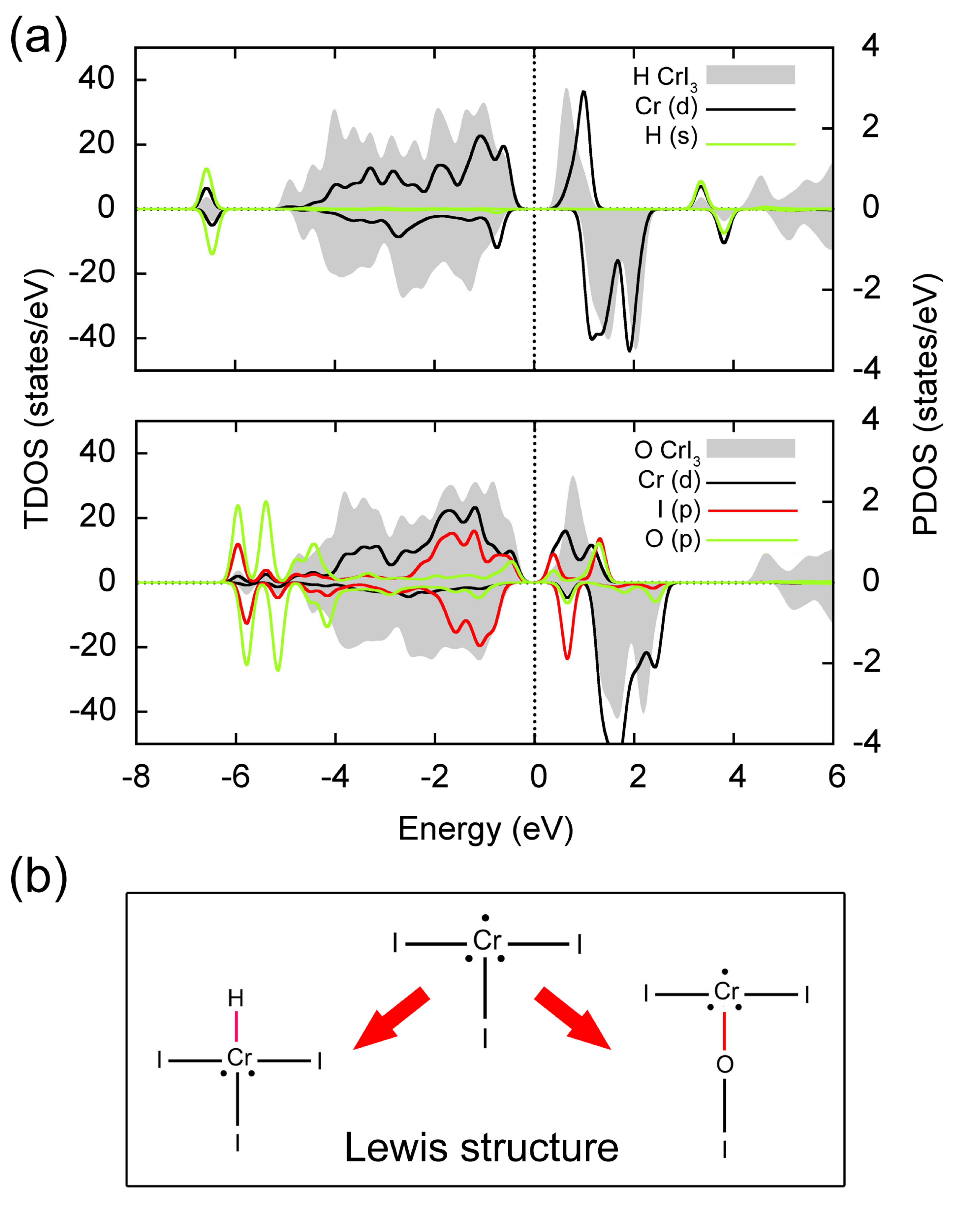}
		\caption{TDOS and PDOS of  HCrI$_3 $ ($ 2\times2 $) system 
			OCrI$_3 $ ($ 2\times2 $) system (a), the positive (negative) values correspond to spin-up (spin-down) states. 
			A schematics of the Lewis structure before and after the H and O atoms adsorption in  CrI$_3 $ (b).}
		\label{fgr:8}
	\end{figure}
	
	On the other hand, the Mulliken population analysis \cite{mulliken1955electronic} yields an 
	excess charge of electrons on H and O adatoms (Fig. \ref{fgr:7}-a). 
	The obtained results of charge transfer are given in Table \ref{tbl:1}. The electronegativity of H, O, Cr 
	and I are 2.20, 3.44, 1.66 and 2.66 respectively. When a Cr atom is exposed to H atom, due to the higher
	electronegativity of H, charge is transferred from Cr to H. In the case of O, the amount of charge transfer is more, due 
	to its even higher electronegativity. 
	%iSo, adsorption of O atoms, would render a p-doped or hole doping CrI$_3 $ monolayer.
	
\begin{figure}[t!]
		\centering
		\includegraphics[width=8.3cm]{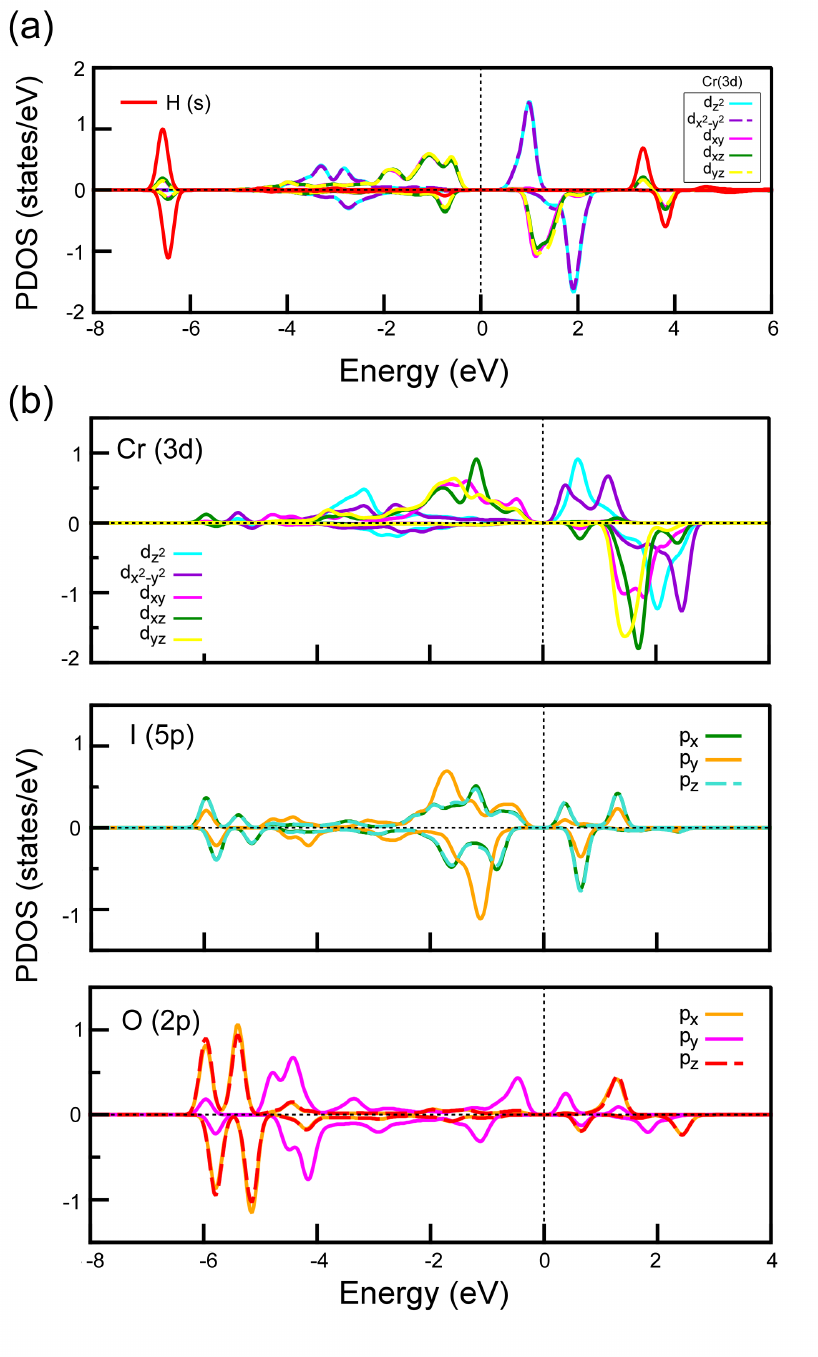}
		\caption{DOS projected onto Cr-d and H-s orbitals for HCrI$ _3 $ system (a), and onto the Cr-d , I-p, and O-p orbitals for OCrI$ _3 $ system (b).}
		\label{fgr:9}
	\end{figure}

	The DOS projected on different atomic species for the  HCrI$_3 $ ($ 2\times2 $) system is presented in Fig. \ref{fgr:8}-a. 
	(The TDOS is for comparative purposes.) Clearly, we can see the hybridization between the 
	1s state of H atom and the 3d states of 
	the nearest Cr atom. For more details we have also calculated the orbital projected PDOS for Cr 3d orbitals 
	(Fig. \ref{fgr:9}-a).
	
	The down-spin $ t_{2g} $ d orbitals are now partially filled, but the octahedral splitting is preserved. As we mentioned, 
	the $ t_{2g} $ levels of Cr in an octahedral environment have 3 unpaired electrons, totalling  $ 3\mu_B $ per atom.
	As a result, when Cr binds to H, it shares one of its d-electrons (one of the three unpaired d electrons pairs with 
	the H s orbital electron) and the total magnetization decreases by 
	$ 1\mu_B $. In brief, due to the reduction of a valence electron per H atom, the overall magnetization is expected to decrease upon H concentration, but probably not completely since it is difficult to imagine a total passivation of Cr atoms by 3 H atoms each.
	
	O and I electronic configurations are $ 2s^2\ 2p^4 $ and $ 5s^2\ 5p^5 $, respectively. Therefore, based on Hund's rule, 
	O and I have 2 and 1 unpaired electrons in their p orbitals, respectively. When the O atom is exposed to the lower I
	atom in the  CrI$_3 $ single-layer, it helps the I atom to break the bonds with the Cr atoms while making new bonds with the displaced I atom and with the Cr atoms by itself. As a result, it shares its two unpaired electrons, and the entire system magnetization remains unchanged. Fig. \ref{fgr:8}-b illustrates the schematic Lewis structure that occurs. The PDOSs of the  OCrI$_3 $ ($ 2\times2 $) system is shown in Fig. \ref{fgr:8}-a. The O p states hybridize with Cr d and I p states (as they show some identical peaks). To shed more light onto the details of this hybridization, we have calculated the orbital projected DOS in this case too (Fig. \ref{fgr:9}-b). Oxygen has completely destroyed the octahedral symmetry. The d states are not degenerate in two groups anymore. In the case of O and I, the main hybridization is between $ p_x/p_z $ orbitals of the two atoms.
	
	Having established the electronic changes induced by the adsorbed species, the physical and measurable consequences are now addressed.
	One of the key measurable quantities that characterizes a ferromagnetic material is its Curie temperature ($ T_C $) at which the 
	ferromagnetic-paramagnetic 
	transition occurs. A common way to study this transition is by mapping our DFT Hamiltonian to a Heisenberg spin Hamiltonian \cite{smart1966effective} from which a single atom Hamiltonian can be written as:
	\begin{equation}
	\mathcal{H}_1=-2\boldsymbol{S}_i.\sum_j{J_{ij}\boldsymbol{S}_j},
	\end{equation}
	where \textbf{S} represents the net magnetic moment at the $ N_i $ site, i and j stand for the nearest $ N_i $ atoms with exchange-coupling parameter J. Replacing the interaction with other spins by an effective magnetic field $ \textbf{H}_{eff} $ one can obtains:
	\begin{equation}
	\mathcal{H}_1=-\vec{\mu}_i.\boldsymbol{H}_{eff}=-g\mu_B\boldsymbol{S}_i.\boldsymbol{H}_{eff},
	\end{equation}
	where $g$ is the Land\'e factor, $ \mu_B $ is the Boltzmann constant, and
	\begin{equation}
	\boldsymbol{H}_{eff}=\frac{2}{g\mu_B}\sum_j{J_{ij}\boldsymbol{S}_j}
	\end{equation}
	Based on the Weiss approximation, each $ \textbf{S}_j $ can be replaced by its average value $ \left\langle\textbf{S}_j\right\rangle=\left\langle\textbf{S}\right\rangle $ 
	\begin{equation}
	\boldsymbol{H}_{eff}=\frac{2\left\langle\textbf{S}\right\rangle}{g\mu_B}\left(\sum_j{J_{ij}}\right)
	\end{equation}
	By assumption, all magnetic atoms are identical and equivalent, this implies that $ \left\langle{\textbf{S}_i}\right\rangle $ is related to the magnetization of the crystal by $ \textbf{M}=ng\mu_B\left\langle{\textbf{S}}\right\rangle $ where n is the number of spins per unit volume.
	\begin{equation}
	\boldsymbol{H}_{eff}=\frac{2}{ng^2\mu^2_B}\left(\sum_j{J_{ij}}\right)\boldsymbol{M}=\lambda\boldsymbol{M}
	\end{equation}
	where $ \lambda $ is the Weiss molecular field coefficient. If there are only nearest-neighbor interaction, $ \sum_j{J_{ij}}=zJ $ where z is the number of nearest-neighbors. Then
	\begin{equation}
	\lambda=\frac{2zJ}{ng^2\mu^2_B}
	\end{equation}
	If there is an external field $ \textbf{H}_{ext} $, then the total field acting on the $ i $th spin is $ \textbf{H} = \textbf{H}_{ext} + \textbf{H}_{eff} $. Having $ M=\chi_C H $ where $ \chi_C $ is Curie susceptibility,
	\begin{equation}
	M=\chi_C(\textbf{H}_{ext} + \textbf{H}_{eff})=\chi_C\textbf{H}_{ext}+\chi_C\lambda \textbf{M}
	\end{equation}
	since $ \chi={\partial M}/{\partial H} $, 
	\begin{equation}
	\chi=\frac{\chi_C}{1-\lambda\chi_C}
	\end{equation}
	putting $ \chi_C=C/T $, where C is Curie constant ($ C={n (g\mu_B)^2 S(S+1)}/{3K_B} $) \cite{reif2009fundamentals}, gives the Curie-Weiss Law:
	\begin{equation}
	\chi(T)=\frac{C}{T-T_C}
	\end{equation}
	where $ T_C=\lambda C $. By substituting $ \lambda $ and C in this equation
	\begin{equation}
	T_C^{\rm MFT}=\lambda C=\frac{2zJS(S+1)}{3k_B}
	\end{equation}
	Thus, based on the Weiss molecular-field theory (MFT), the $ T_C $ can be simply estimated using the 
	nearest-neighbor exchange-coupling parameter $J$, being $ z=3 $  the number of  nearest-neighbor Cr atoms in the  CrI$_3 $ monolayer, $ S=3/2 $, and $ k_B $ is 
	the Boltzmann constant \cite{he2017near}. 
	
	We have computed  exchange coupling constants using  $ jx $, a post-processing code for OpenMX , \cite{han2004electronic,terasawa2019efficient}. We have first computed  $J$ for pristine CrI$_3$, obtaining the nearest-neighbor value and corresponding Curie temperature that are presented in Table \ref{tbl:3}. Considering now the H adatom case, we have computed the Cr-Cr nearest-neighbor coupling constants between all Cr atoms and taken the average, obtaining a higher value and a higher Curie temperature, as indicated in Table \ref{tbl:3}. Repeating this procedure for the case of O, we  obtain, however,  a smaller Curie temperature of 55 K.  Therefore, the adsorption of both species, even at low coverage, translates into a significant 
	change of the Curie temperature with respect to that of pristine  CrI$_3 $.
	
	\begin{table}[b]
		\small
		\caption{\ The (average) nearest-neighbor exchange-coupling parameter $J$ and  the Curie temperature ($ T_C $) based on the Weiss 
			molecular-field theory using the $jx$ code and Ising model to compute the coupling constants , along with the rescaled absolute values.}
		\label{tbl:3}
		\begin{tabular*}{0.48\textwidth}{@{\extracolsep{\fill}}cccc}
			\hline
			&  CrI$_3 $ &  HCrI$_3 $ & 	 OCrI$_3 $ \\
			\hline
			$ J ({\rm meV}) {\rm (jx)} $ & 0.70 & 0.90 & 0.64 \\
			$ T_C (^{\circ}K)\ {\rm (jx)} $ & 61 & 78 & 55 \\
			$ T_C (^{\circ}K)\ {\rm (jx/rescaled)} $ & - & 57 & 41 \\
			$ J ({\rm meV}) \ {\rm (Ising)}  $ & 0.95 & 1.02 & 1.06  \\
			$ T_C (^{\circ}K)\ {\rm (Ising)} $ & 83 & 89 & 92 \\
			$ T_C (^{\circ}K)\ {\rm (Ising/rescaled)} $ & - & 48 & 50 \\
			\hline
		\end{tabular*}
	\end{table}
	
	The MFT $ T_C $ for the H/O functionalized  CrI$_3 $ monolayer can be corrected by an appropriate rescaling \cite{liu2018screening,he2019remarkably}
	\begin{equation}
	\frac{T_C\rm{(XCrI_3)}}{T_C^{\rm MFT}\rm{(XCrI_3)}}=\frac{T_C^{\rm exp}{\rm (CrI_3)}}{T_C^{\rm MFT}{\rm (CrI_3)}}
	\end{equation}
	where $X$  represents H or O, $ T_C{\rm(XCrI_3)} $ is the rescaled value, $ T_C^{\rm exp}{\rm (CrI_3)} $ is 
	the experimental one for the clean monolayer (45K), 
	and $ T_C^{MFT}{\rm (CrI_3)} $ represents the Curie temperature calculated by mean field theory (MFT). Even after this 
	rescaling, the increase in the Curie temperature compared with that of pristine  CrI$_3 $ is noticeable.
	
	Notice, however, that these calculations contain some approximations in addition to the ones already inherent in  MFT.  The $J$ coupling constants are assumed to be equal in the MFT approximation, but the  computed  ones  are not  constant across the cell, with fluctuations of up  to 100\%. Furthermore,  spin-orbit coupling (SOC) has not been considered  and is expected to increase the Curie temperature due to anisotropy contributions. Therefore, we now turn to another method. We  map the DFT + SOC results for the fully ferromagnetic  and fully antiferromagentic ground state energies to those of an Ising Hamiltonian:  
	\begin{equation}
	\mathcal{H_I}=-\sum_{i,j}{J_{ij}\boldsymbol{I}_i\boldsymbol{I}_j}.
	\end{equation}
	We obtain a value of $J=0.95$ meV for pristine CrI$_3$,  which  translates into a Curie temperature of 83 K. These numbers compare rather well with those obtained with the $jx$ code and, as expected, are slightly higher. The results for the H and O adatom cases are also presented in Table \ref{tbl:3}. We see that in both cases the Curie temperature increases now.  We attribute the discrepancy in the case of  O to the fact that the fluctuations of the $J$ coupling constants are visibly larger than for H, and the MFT is probably less reliable.
	
\begin{figure}[t!]
		\centering
		\includegraphics[width=8.3cm]{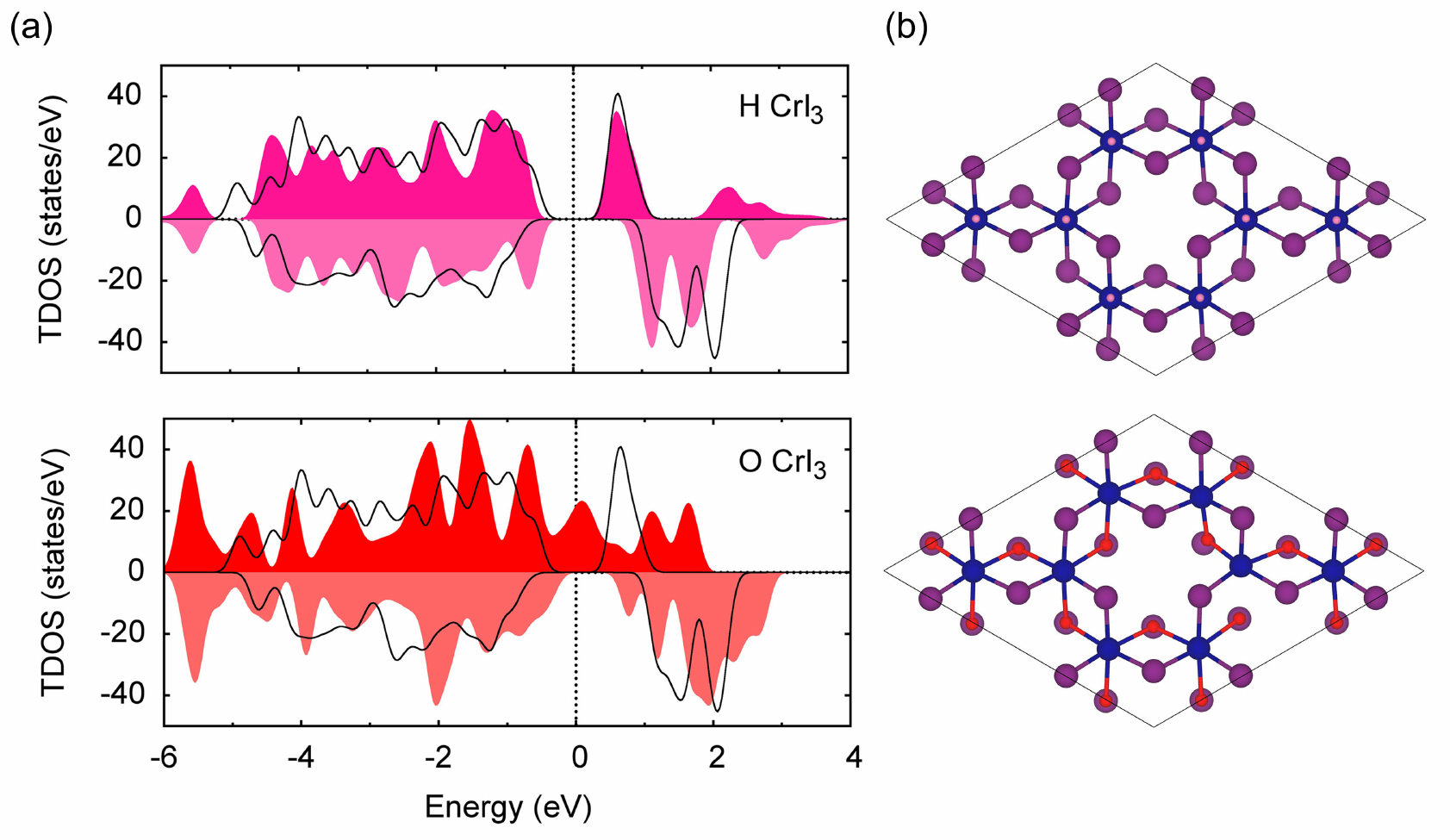}
		\caption{TDOS of fully functionalized HCrI$_3 $ and  OCrI$_3 $ systems. The positive (negative) values correspond to spin-up (spin-down) states. Black lines show The TDOS of the pure  CrI$_3 $ (a), and fully functionalized CrI$ _3 $ ($ 2\times2 $) structures with Hydrogen and Oxygen atoms (b).}
		\label{fgr:10}
	\end{figure}

	Finally, we fully functionalize the CrI$ _3 $ surface with H and O atoms. Since the favorable site for the H atom was found to be on top of the Cr atom, here all the Cr atoms have H atoms on  top. In the case of O, the  atoms are placed on V$ _I $ sites, as this is the most stable configuration. The TDOS of the fully functionalized HCrI$ _3 $ and OCrI$ _3 $ systems is plotted in Fig.  \ref{fgr:10}-a along with that of the clean CrI$ _3 $ for comparative purposes. Here again, as we can see in this figure, H leaves the TDOS of CrI$ _3 $ almost without change, while O introduces visible changes.
	The full functionalization of the CrI$ _3 $ system with O atoms has changed its electronic structure into that of a half-metal. This means that the spin-up channel crosses the Fermi level, while we have a gap in the opposite channel. Therefore, it acts as a conductor to spin-up electrons but as a semiconductor to those of the spin-down electrons. The total magnetization of the system decreases to 16$ \mu_B $ in case of HCrI$ _3 $ (i.e. its total magnetism decreases by one per H atom) while it remains unchanged for fully functionalized OCrI$ _3 $ system. 
	
	The increase of the Curie temperature has been confirmed with two different calculations for one H adatom, so we expect full coverage to have a stronger impact. The Curie temperature can now be safely computed with the "jx" code since all Cr atoms are identical and, expectedly, we see a very significant increase: $ T_C^{\rm jx}= 192^{\circ}K $ and $ T_C \ (rescaled)= 142^{\circ}K $. We have ignored SOC again here, although, as we have seen, it may reflect in a relatively lower increase of the Curie temperature. This calculation confirms that hydrogenation can serve as a simple mechanism to enhance the Curie temperature. In addition,  the main properties of CrI$_3 $ (such as its ferromagnetic feature and semiconductor band structure) remain unchanged. 
	
	\section{Conclusions}
	We have investigated, through DFT calculations, the electronic and magnetic structures of the 2D monolayer 
	FM insulator  CrI$_3 $ resulting from H and O adsorption.  We have found that 
	H adatoms preferentially adsorb on Cr atoms, but hardly influence the original  CrI$_3 $ structure although 
	reduce the magnetic moment of the hosting Cr atoms.
	Oxygen adatoms present a much higher adsorption energy (also on Cr atoms) and greatly distort the atomic structure while
	introducing an impurity band in the gap.
	Interestingly, we have predicted that the intrinsic ferromagnetism and corresponding Curie temperature can be 
	largely enhanced by the adsorption of H  atoms up to  approx. 200 \% for full H coverage. Since the main properties of the system remain unchanged, hydrogenation may be a viable route towards room temperature operation devices. Oxygen adsorption, on the other hand, offers the possibility of turning CrI$_3$ into a half metal.
	
	\section{Acknowledgements}
	JJP acknowledges Spanish MINECO through Grant FIS2016-80434-P, the Fundaci\'on Ram\'on Areces, the Mar\' ia de Maeztu 
	Program for Units of Excellence in R\&D (MDM-2014-0377), the Comunidad Auto\'onoma de Madrid through S2018/NMT-4321 
	(NanomagCOST-CM) and the European Union Seventh Framework Programme under Grant agreement No. 604391 Graphene Flagship. 
	JJP also acknowledges the computer resources and assistance provided by the Centro de Computación Científica of the 
	Universidad Autónoma de Madrid and the RES.
	
	\bibliography{apssamp}% Produces the bibliography via BibTeX.
	
\end{document}